\documentclass[sn-mathphys-num]{sn-jnl}% Math and Physical Sciences Numbered Reference Style 
\usepackage{graphicx}%
\usepackage{multirow}%
\usepackage{amsmath,amssymb,amsfonts}%
\usepackage{amsthm}%
\usepackage{mathrsfs}%
\usepackage[title]{appendix}%
\usepackage{xcolor}%
\usepackage{textcomp}%
\usepackage{manyfoot}%
\usepackage{booktabs}%
\usepackage{algorithm}%
\usepackage{algorithmicx}%
\usepackage{algpseudocode}%
\usepackage{listings}%
\def\ini{\mathrm{ini}}
\def\msun{M_{\odot}}
\def\Msun{$~M_{\odot}$}

\theoremstyle{thmstyleone}%
\theoremstyle{thmstyletwo}%

\theoremstyle{thmstylethree}%

\begin{document}

%\title[Article Title]{Nucleosynthesis in AGB stars: from Lithium to Uranium}
%\title[Article Title]{Synthesis of actinides and short-lived radionuclides in AGB stars experiencing proton ingestion}
\title[Article Title]{Synthesis of actinides and short-lived radionuclides during i-process nucleosynthesis in AGB stars }
%\title[Article Title]{Synthesis of thorium, uranium and short-lived radionuclides in AGB stars experiencing proton ingestion}


\author*[1]{\fnm{Arthur} \sur{Choplin}}\email{arthur.choplin@ulb.be}

\author[1]{\fnm{Stephane} \sur{Goriely}}%\email{iiauthor@gmail.com}

\author[1]{\fnm{Lionel} \sur{Siess}}%\email{iiiauthor@gmail.com}

\author[1]{\fnm{S\'ebastien} \sur{Martinet}}%\email{iiiauthor@gmail.com}

%; arthur.choplin@ulb.be

\affil*[1]{Institut d'Astronomie et d'Astrophysique, Universit\'e Libre de Bruxelles,  CP 226, B-1050 Brussels, Belgium}


\abstract{\textbf{Context:} A complex interplay between mixing and nucleosynthesis is at work in asymptotic giant branch (AGB) stars. In addition to the slow neutron capture process (s-process), the intermediate neutron capture process (i-process) can develop during protons ingestion events (PIEs). 

\textbf{Purpose:} In this paper, after quickly reviewing the different modes of production of heavy elements in AGB stars that were identified so far, we investigate the synthesis of actinides and other short-lived radioactive nuclei (SLRs, $^{60}$Fe, $^{107}$Pd, $^{126}$Sn, $^{129}$I, $^{135}$Cs and $^{182}$Hf) during i-process nucleosynthesis.

\textbf{Methods:} AGB stellar models with initial masses $1 \leq M_\ini/\msun \leq 3$, metallicities $-3 \leq $~[Fe/H]~$ \leq 0$  and different overshoot strengths were computed with the stellar evolution code \textsf{STAREVOL}. During PIEs, a nuclear network of 1160 isotopes is used and coupled to the transport equations.

\textbf{Results:} We found that AGB models with [Fe/H]~$<-2$ can synthesize actinides with sometimes abundances greater than solar. The $^{60}$Fe yield scales with the initial metallicity while the $^{107}$Pd, $^{126}$Sn, $^{129}$I, $^{135}$Cs and $^{182}$Hf yields follow a similar pattern as a function of metallicity, with a production peak at [Fe/H]~$\simeq -1.3$.
At [Fe/H]~$<-1$, the fraction of odd Ba isotopes $f_{\rm Ba,odd}$ is predicted to vary between 0.6 and 0.8 depending on the initial mass and metallicity. Nuclear uncertainties on our 1\Msun\, [Fe/H]~$=-2.5$ model lead to $f_{\rm Ba,odd}$ ranging between 0.27 and 0.76, which is clearly above the s-process value.

\textbf{Conclusion:} AGB stars experiencing PIEs appear to be potential producers of actinides and SLRs, particularly at low metallicity (except for $^{60}$Fe). Galactic chemical evolution modeling are required to assess their possible contribution to the galactic enrichment. 
 }

%%================================%%
%% Sample for structured abstract %%
%%================================%%




\maketitle

\section{Introduction}\label{sec1}

Before crossing the planetary nebula phase and dying as white dwarfs, stars with initial masses $0.8<M_\ini/\msun< 8$ go through the asymptotic giant branch (AGB) phase \citep{busso99, herwig05, karakas14}. AGB stars are made of a carbon-oxygen (CO) core, surmounted by thin He- and H-burning shells.  Thermal instabilities in the helium-burning shell produce recurrent convective thermal pulses (TP). When fully developed, the top of the convective TP can encroach the bottom of the hydrogen-rich layer, leading to a proton ingestion episode (PIE, e.g. \cite{iwamoto04, campbell08, cristallo09a, suda10, gilpons22, choplin21, remple24}). 
When protons are engulfed in the TP, the reaction chain $^{12}$C($p,\gamma$)$^{13}$N($\beta^{+}$)$^{13}$C($\alpha,n$)$^{16}$O activates, producing neutron densities up to $N_n \simeq 10^{15}$~cm$^{-3}$ and eventually leading to an i-process nucleosynthesis (e.g. \cite{cristallo09a, choplin21}), accompanied by a copious production of $^{7}$Li and $^{13}$C \cite{iwamoto04, choplin24c}. 
Actinides (particularly Th and U) were shown to be significantly synthesized during the PIE of a 1\Msun\ AGB model at [Fe/H]~$=-2.5$ \cite{choplin22b}, although these results strongly depend on nuclear uncertainties \cite{martinet24}. 
Some short-lived radionuclides\footnote{Short-lived radionuclides have half-lives between 0.1 and 100 Myr.} (SLR, e.g. \cite{lugaro18} for a recent review) may also be produced during the i-process nucleosynthesis. For instance, a non-negligible production of the $^{129}$I (half-life of $t_{\rm 1/2} = 16$~Myr) and $^{126}$Sn ($t_{\rm 1/2} = 0.23$~Myr) isotopes require neutron densities above $\sim 10^{12} - 10^{13}$~cm$^{-3}$, which may not be reached during the s-process nucleosynthesis. Up to now, the stellar production of these two SLRs have thus been attributed to the r-process nucleosynthesis \cite{meyer00, lugaro18}. During the i-process nucleosynthesis, however, neutron densities are high enough to synthesize such isotopes. 

In this paper, we investigate the production of actinides and six SLR ($^{60}$Fe, $^{107}$Pd, $^{126}$Sn, $^{129}$I, $^{135}$Cs and $^{182}$Hf) in AGB stars experiencing an i-process nucleosynthesis over a wide range of masses and metallicities.

\begin{figure}[t]
\centering
\includegraphics[scale= 0.27,  trim={10cm 2cm 5cm 5cm},clip]{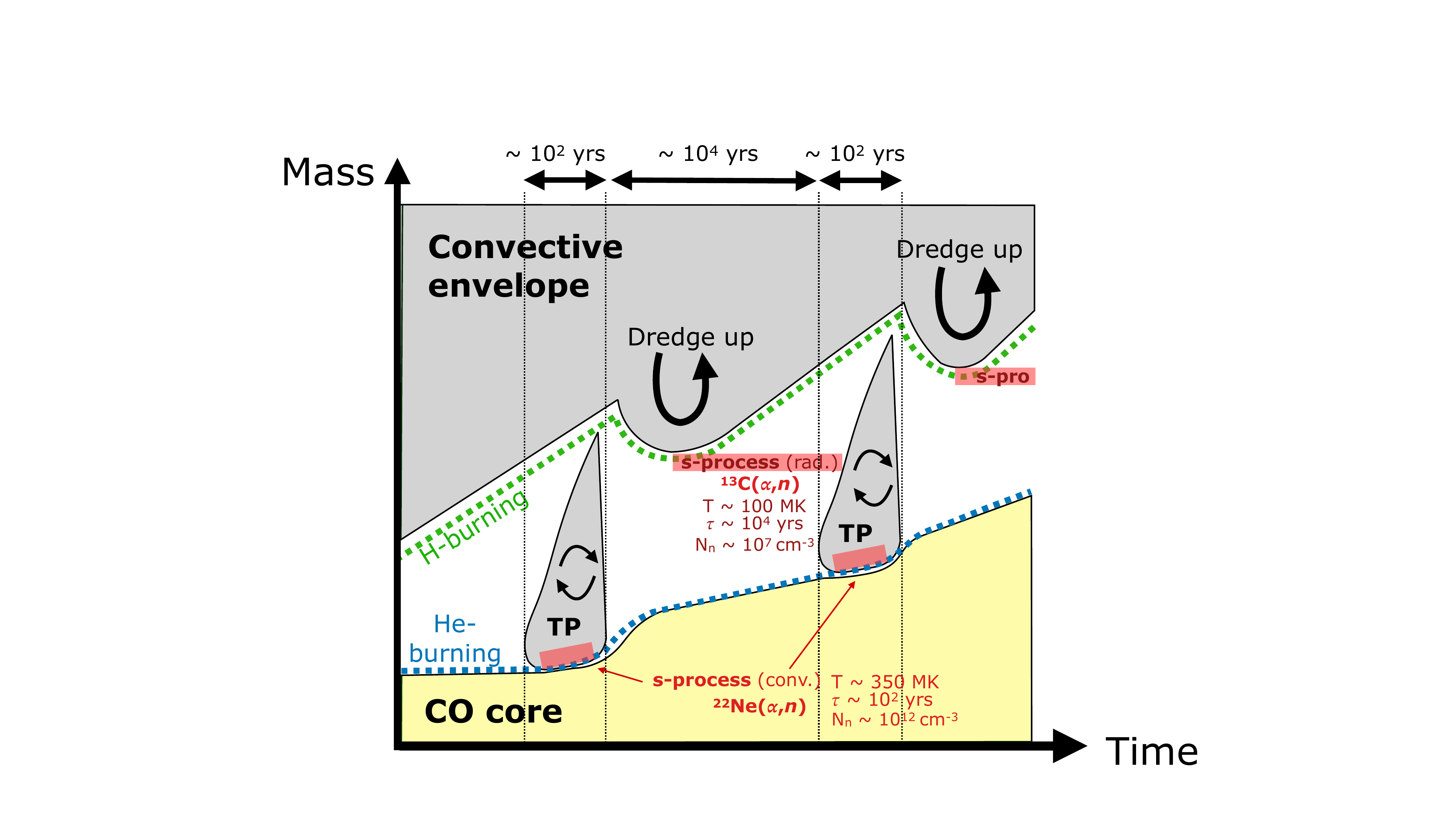}
\caption{Schematic view of the internal structure of an AGB star without PIE. Grey zones show convective zones. The yellow area corresponds to the CO-core. The green (blue) dashed lines represent the location of the H-burning (He-burning) shell. The red shaded areas show where the radiative and convective s-processes operate. }
\label{fig:agb1}
\end{figure}

\begin{figure}[t]
\centering
\includegraphics[scale= 0.27,  trim={10cm 2cm 0cm 5cm},clip]{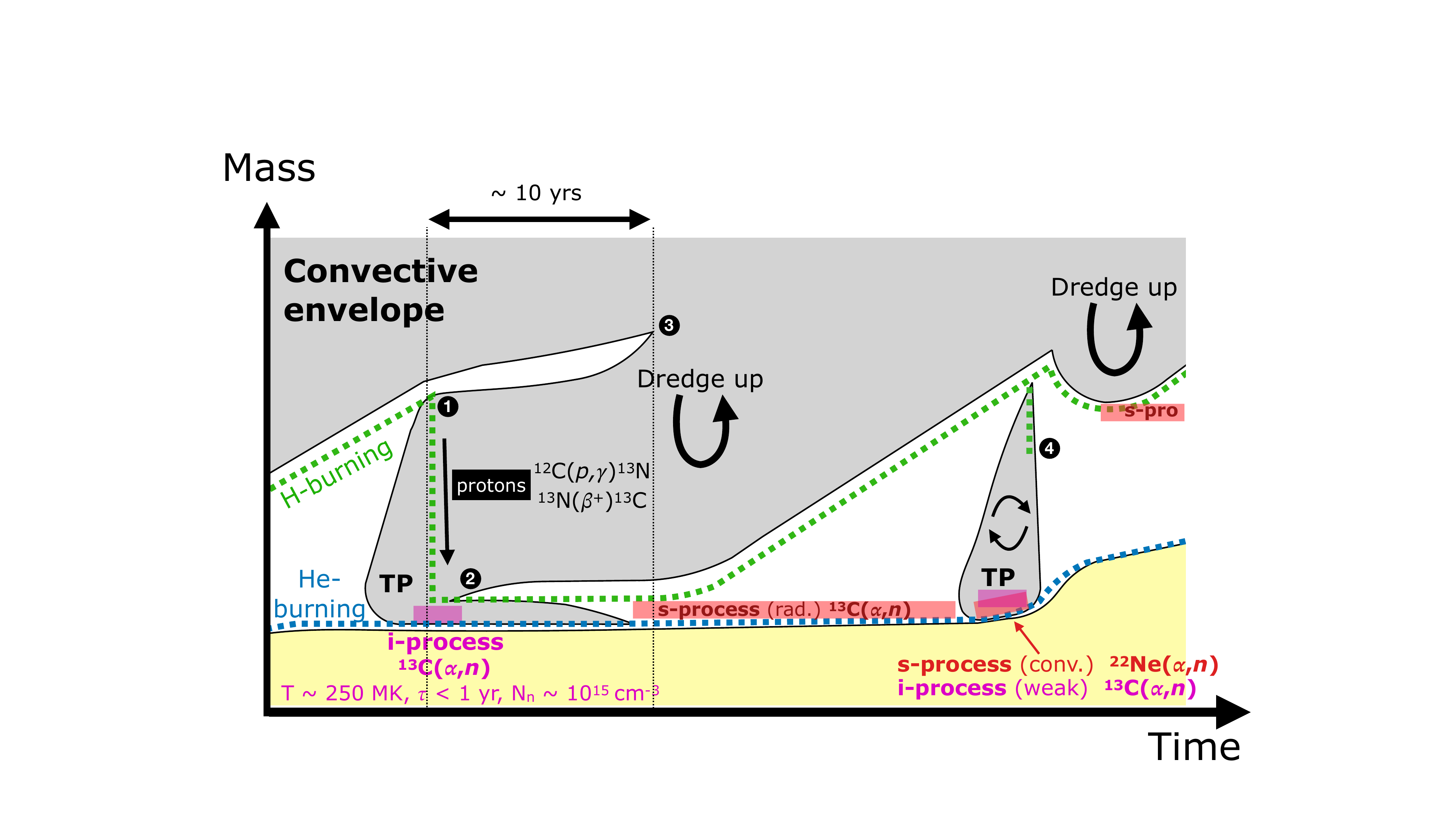}
\caption{Schematic view of the internal structure of an AGB star that experienced a PIE. Grey zones show convective zones. The yellow area corresponds to the CO-core. The green (blue) dashed lines show where the energy generated by H-burning (He-burning) is maximal. The red (magenta) shaded areas show where the s-process (i-process) operates. The numbers indicate different important events : (1) start of PIE, (2) split of the TP, (3) merging of pulse with envelope and (4) weak PIE (cf. text for more details).}
\label{fig:agb2}
\end{figure}

%%%%%%%%%%%%%%%%%%%%%%%%%%%%%%%%%%%%%%%%%%
\section{AGB models}\label{sec2}

The AGB models considered in this paper were presented in \cite{choplin22a, choplin24a, choplin24b}. 
The most important aspects are recalled here and we refer the reader to the aforementioned works for further details. 
The models were computed with the stellar evolution code \textsf{STAREVOL} \citep{siess00, siess06, goriely18c}, with initial masses $1 \leq M_\ini/\msun \leq 3$ and metallicities $-3 \leq $~[Fe/H]~$ \leq 0$, using the solar mixture of \cite{asplund09}. 
To follow the i-process nucleosynthesis, a nuclear network of 1160 nuclei is used, linked by 2123 nuclear reactions ($n$-, $p$-, $\alpha$-captures and $\alpha$-decays), weak (electron captures, $\beta$-decays), and electromagnetic interactions. Nuclear reaction rates were taken from the Nuclear Astrophysics Library of the Université Libre de Bruxelles BRUSLIB\footnote{http://www.astro.ulb.ac.be/bruslib/} \cite{arnould06} and the NETGEN interface \cite{xu13} (we refer to \cite{choplin21,goriely21,choplin22a} and references therein for additional details). During a PIE, the convective and nuclear burning timescales become similar imposing the nucleosynthesis and transport of chemical species to be solved simultaneously (see Sect. 2.1 in \cite{choplin22a} for more details). 
In a number of our models, overshooting above the TP was included following the exponential law introduced in \cite{goriely18c}. The extent of the overshoot mixing zone beyond the Schwarzschild boundary is controlled by the parameter $f_{\rm top}$ \citep[cf. Sect.~2 in][]{choplin24a}, the value of which is varied in the range $f_{\rm top} = 0.02$, $0.04$, $0.10$ or $0.12$. Overshoot mixing above the TP was shown to favour the occurrence of PIEs \cite{choplin24a, remple24}. Without extra mixing, PIEs develop in AGB stars having $M_\ini \lesssim 2.3$\Msun\ and [Fe/H]~$\lesssim -2$. With $f_{\rm top} = 0.10$, PIEs develop up to [Fe/H]~$= -0.5$ if $M_\ini < 3$\Msun \cite{choplin24a}.

\begin{figure}[t]
\centering
\includegraphics[width=0.7\textwidth]{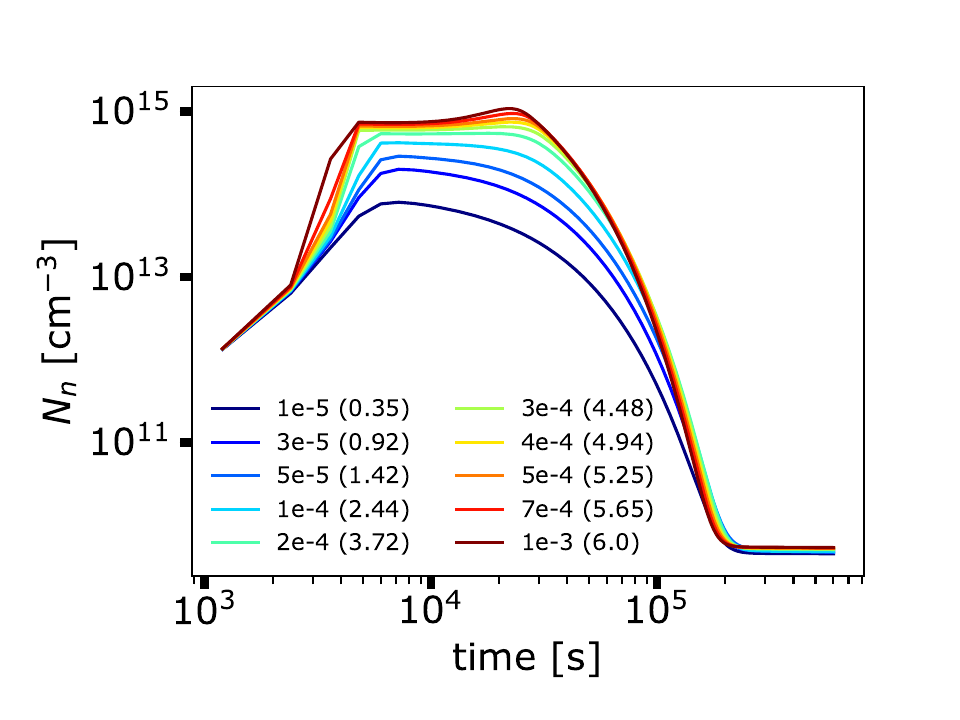}
\caption{Neutron density $N_n$ as a function of time for one-zone calculations with $T=220$~MK, $\rho = 2500$~g cm$^{-3}$ and initial proton mass fractions between $10^{-5}$ and  $10^{-3}$ (labels of the lines). The number in parenthesis indicates the total neutron exposure $\tau_{\rm exp} = \int_{t_1}^{t_2}  N_{\rm n}(t) \, v_{\rm T}(t) \, \text{d} t$ in mbarn$^{-1}$ where $v_{\rm T} = \sqrt{ \, 2 \, k_B \, T(t) /m_{\rm n}}$ is the neutron thermal velocity with $k_B$ the Boltzmann constant, $T(t)$ the temperature at time $t$ and $m_n$ the neutron mass.}
\label{fig:onezone_nn}
\end{figure}

\subsection{Three modes of production of heavy elements in AGB stars}

Figure~\ref{fig:agb1} shows a schematic view of a thermally pulsating AGB star model without PIE. During the interpulse period and if some extra mixing (e.g. overshoot) is present below the convective envelope at the time of the third dredge up, protons can diffuse in the $^{12}$C-rich layers left by the TP. This leads to the formation of a $^{13}$C-pocket and the production of trans-iron elements via the radiative s-process. Typical temperature, timescale and neutron density during the radiative s-process are $T = 100$~MK, $\tau = 10^{4}$~yr and $N_n = 10^7$~cm$^{-3}$ respectively. The convective s-process can develop at the bottom of the convective TP if the temperature exceeds $T \simeq 350$~MK, which is the case in AGB stars with $M_{\rm ini} \gtrsim 3 M_{\odot}$. The s-process products are later transported to the surface by third dredge up events (e.g. \cite{busso99, herwig05, karakas14} for reviews). 

As first discussed in \citet{schwarzschild67}, PIE can develop during helium shell flashes. Figure~\ref{fig:agb2} shows a schematic (yet realistic) view of a model experiencing a PIE at the beginning of the AGB phase. Quickly after the start of the PIE, the convective pulse splits (point 2 in Fig.~\ref{fig:agb2}) where the timescale associated with the reaction $^{12}$C($p,\gamma$) becomes similar to the convective timescale (see e.g. Sect.~3.5 in \cite{choplin22a} for more details). About 10 yr later, the convective pulse merges with the envelope (point 3 in Fig.~\ref{fig:agb2}) resulting in a strong alteration of the AGB surface chemical composition and subsequent evolution. In particular, the change in the envelope metallicity and opacity boost the mass loss rate. In AGB with $M_{\rm ini} \lesssim 1.5$\Msun, the entire convective envelope is lost before any new TP develops. In more massive AGB stars, the thermally pulsating AGB phase resumes and further TPs develop, with possibly further radiative and convective s-process nucleosynthesis episodes (as described above and shown in Fig.~\ref{fig:agb1}). 
A $^{13}$C-pocket naturally emerges from the leftovers of a PIE (long red shaded band in Fig.~\ref{fig:agb2}, see also Fig.~7 in \cite{choplin22a}) and the nucleosynthesis proceeds in conditions similar to the standard radiative s-process. 
After a PIE, the envelope is enriched in metals and the situation resembles that of a star of higher metallicity, where PIEs are less prone to develop (e.g. Sect~3.1.2 in \cite{choplin22a}). In our models, we never obtain a second PIE but sometimes weak PIEs occur (point 4 in Fig.~\ref{fig:agb2}), leading to $N_n \simeq 10^{12} - 10^{13}$~cm$^{-3}$ at maximum and no noticeable modification of the AGB structure.

%%%%%%%%%%%%%%%%%%%%%%%%%%%%%%%%%%%%%%%%%%
\section{Results}\label{sec3}

\begin{figure}[t]
\centering
\includegraphics[width=1\textwidth]{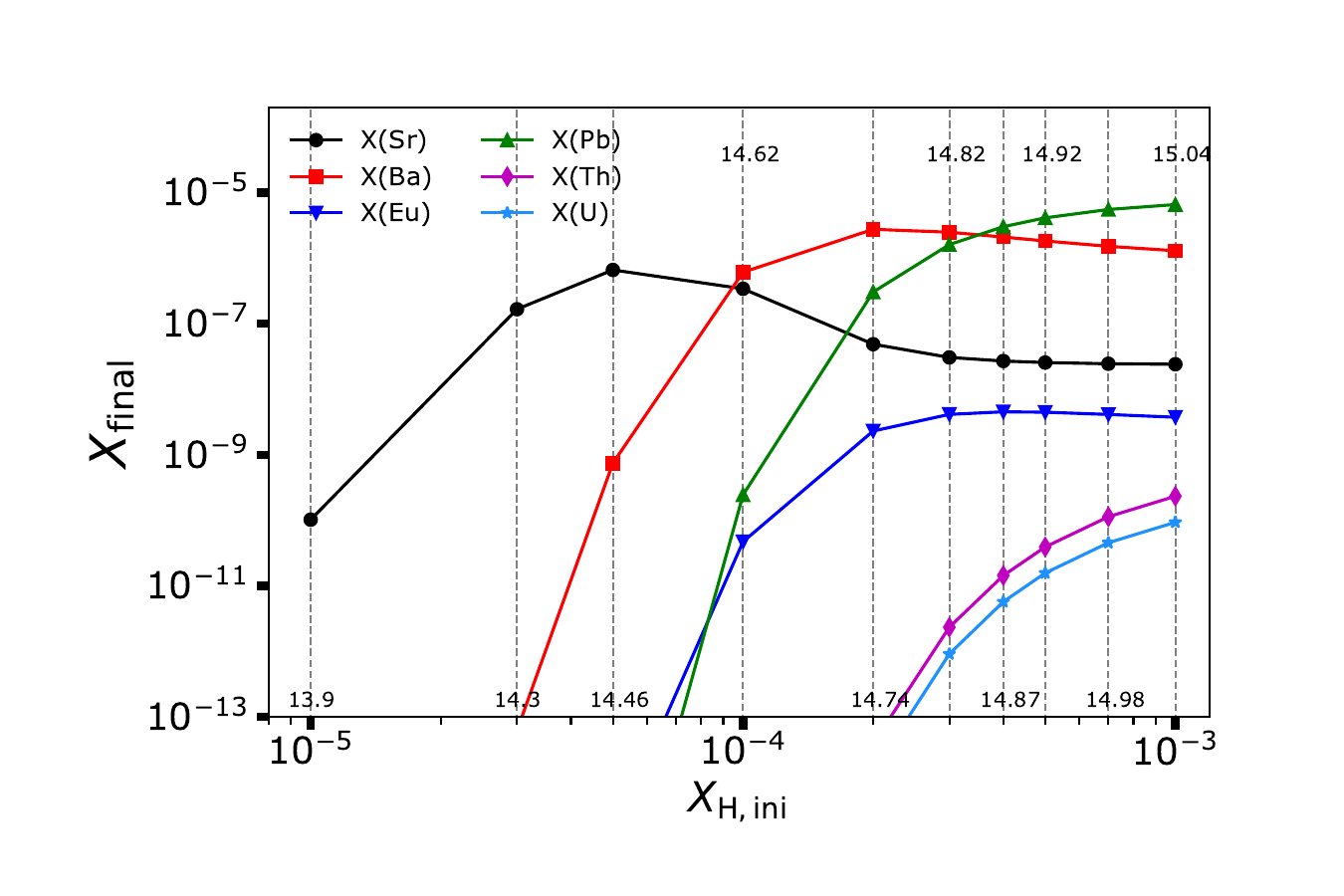}
\vspace{-0.8cm}
\caption{Results obtained from one-zone calculations at metallicity [Fe/H]~$=-2.5$, with $T=220$~MK and $\rho = 2500$~g cm$^{-3}$. The final abundances (after $6 \times 10^5$~s) of 6 elements are shown (after decays of unstable isotopes except  $^{232}$Th  and $^{238}$U, with half lives of 14 and 4.46~Gyr respectively) as a function of the initial proton mass fraction $X_{\rm H, ini}$. The numbers indicate the logarithm of the maximal neutron density reached during each calculation.}
\label{fig:onezone}
\end{figure}

\begin{figure}[t]
\centering
\includegraphics[width=1\textwidth]{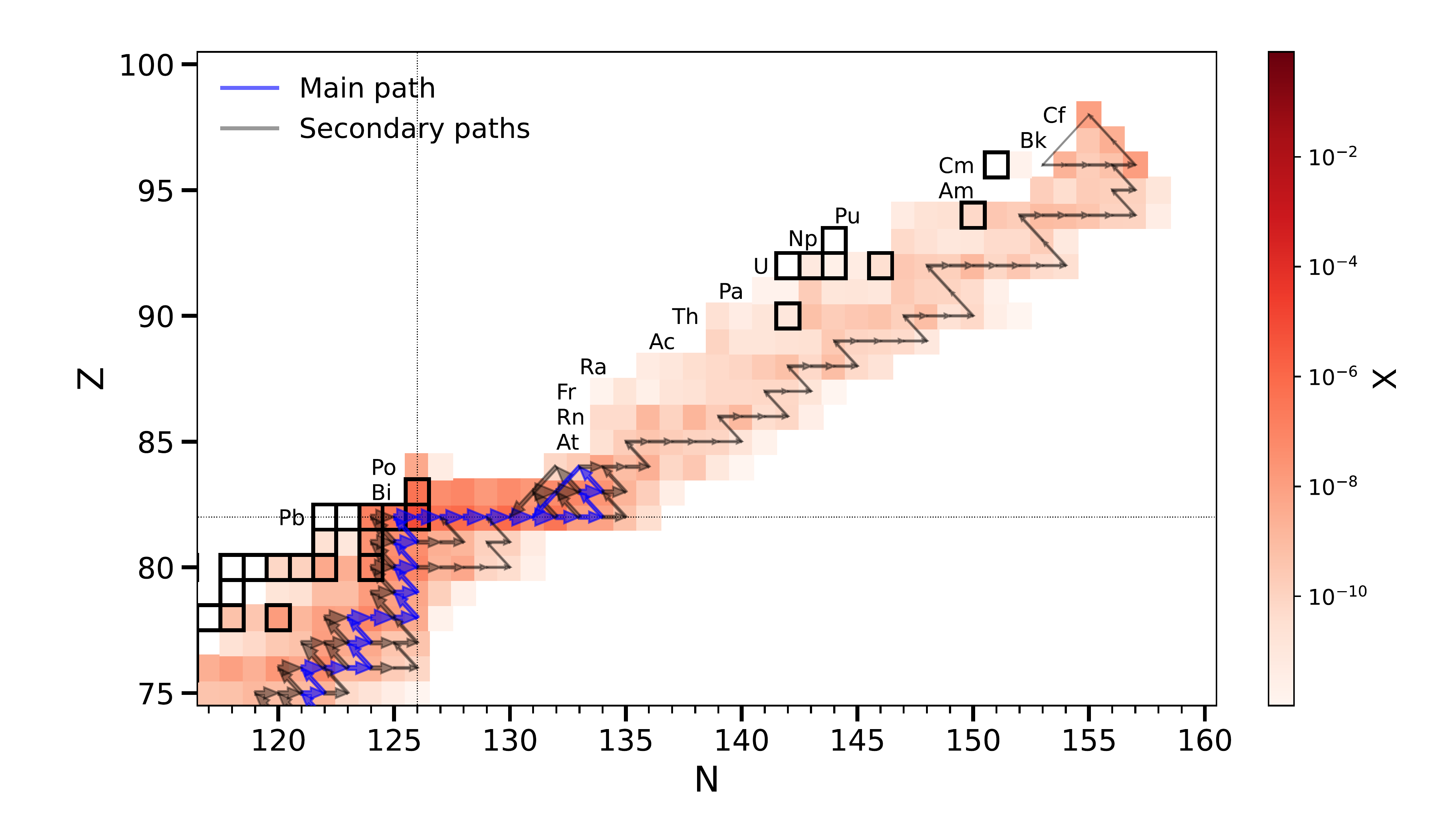}
\caption{Main (blue) and secondary (black) i-process paths (starting from $^{56}$Fe) in a 1~$M_{\odot}$ AGB model at [Fe/H]~$=-2.5$, at the bottom of the convective TP at the time of maximum neutron density. 
A secondary path is considered as such if at least 30~\% of the total flux goes through it. 
The size of the arrows scales with the flux. 
The black squares highlight the stable and long-lived isotopes. 
The colour of the different nuclei corresponds to their mass fraction at that time.
This is the corrected version of the Figure~1 published in \cite{choplin22b} (see text for details). 
}\label{fig:thu}
\end{figure}

\begin{figure}[t]
\centering
\includegraphics[width=0.49\textwidth]{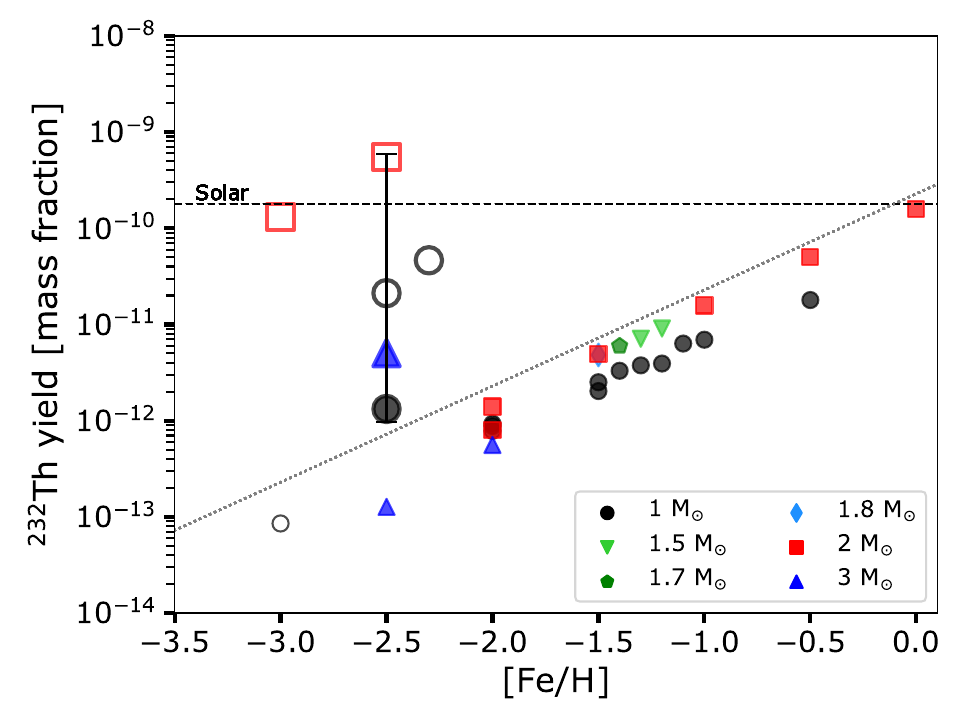}
\includegraphics[width=0.49\textwidth]{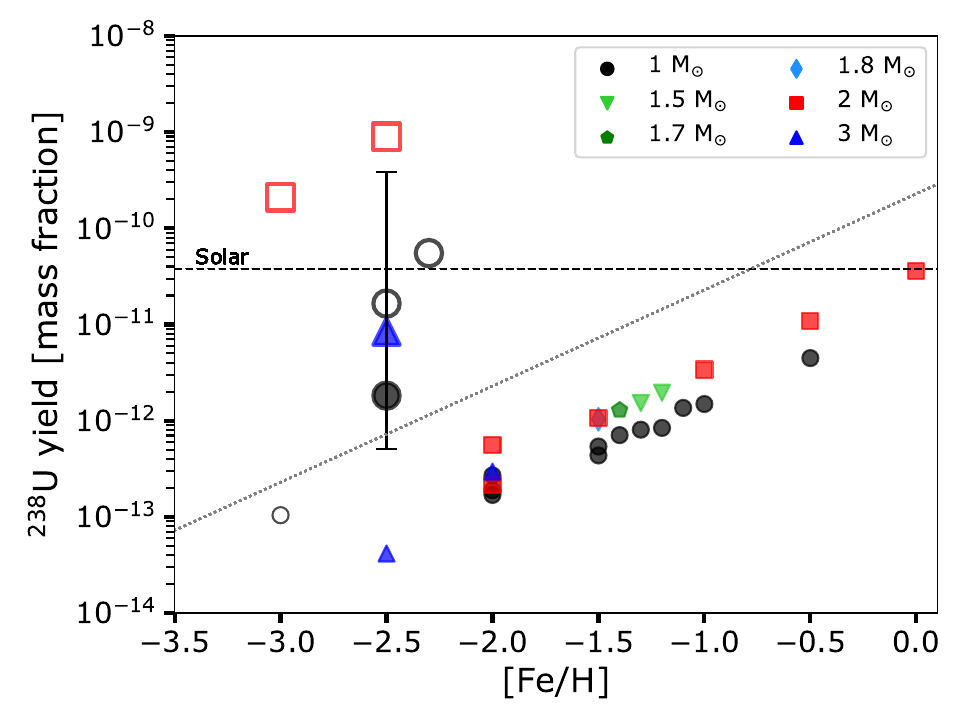}
\caption{Yields in mass fraction of $^{232}$Th and $^{238}$U from AGB models with different initial masses that experienced PIEs, as a function of metallicity. All actinides were decayed except $^{232}$Th ($t_{1/2} = 14$~Gyr) and $^{238}$U ($t_{1/2} = 4.46$~Gyr). The horizontal dashed line shows 
the present-day solar mass fractions of $^{232}$Th and $^{238}$U. 
The oblique dotted line represents the initial sum of actinides in mass fraction in our models (which corresponds to the present-day sum of actinides in the Sun, scaled to the model metallicity). Larger symbols highlight the models above this line, i.e. for which there is an extra production of the considered isotope, coming from Pb and Bi (see text for details). Open (filled) symbols represent models computed without (with) overshooting. Filled symbols with identical shape at a given metallicity (e.g. the two filled blue triangles at [Fe/H]~$=-2.5$) correspond to models of same initial mass computed with different overshooting strengths $f_{\rm top}$ (cf. Sect.~\ref{sec2}). The error bars show nuclear parameter uncertainties associated with our [Fe/H]~$=-2.5$, 1~$M_{\odot}$ AGB model without overshoot, computed in \cite{martinet24} (see Sect.~\ref{sec:agb} for details)}
\label{fig:actiyields1}
\end{figure}

\begin{figure}[h!]
\centering
\includegraphics[width=0.49\textwidth]{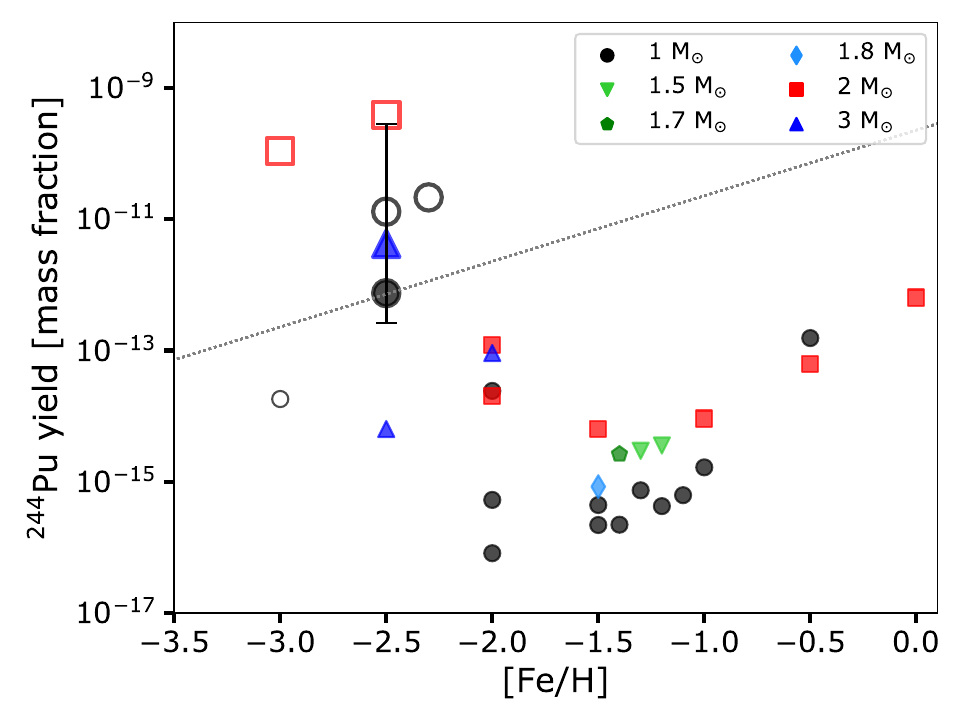}
\includegraphics[width=0.49\textwidth]{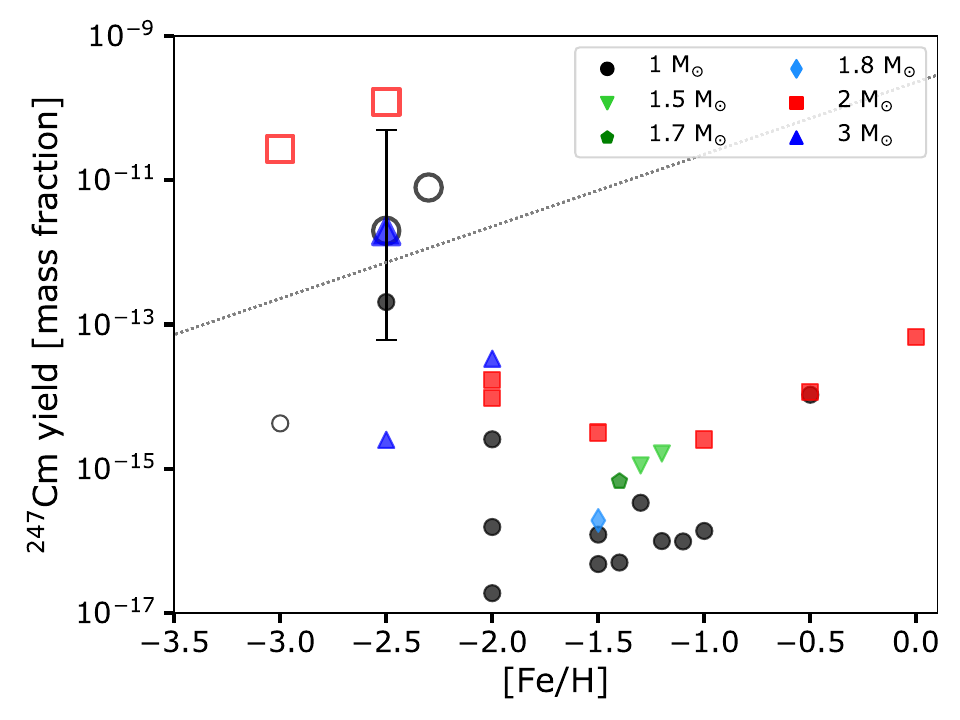}
\caption{Same as Fig.~\ref{fig:actiyields1} but for the SLR $^{244}$Pu ($t_{1/2} = 81.3$~Gyr) and $^{247}$Cm ($t_{1/2} = 15.6$~Gyr). In this case, all actinides were decayed except $^{232}$Th, $^{235}$U, $^{236}$U, $^{238}$U, $^{237}$Np, $^{244}$Pu and $^{247}$Cm, which have $t_{1/2}>1$~Myr. }
\label{fig:actiyields2}
\end{figure}

\subsection{Production of actinides}

The s-process was shown to terminates in a cycle involving Pb, Bi, and Po \cite{clayton67}. When reaching $^{209}$Pb ($t_{1/2} = 3.2$~hr) the flux decays to $^{209}$Bi, 
goes to $^{210}$Bi ($t_{1/2} = 5.0$, $3.7$ and $0.5$~days in terrestrial conditions, $T=100$ and 250~MK, respectively), decays to $^{210}$Po ($t_{1/2} = 138$~days) and finally $\alpha$-decays to the stable $^{206}$Pb. 
This cycle can be broken if the $^{209}$Pb $\beta^-$ decay is bypassed by neutron captures. At $T=250$~MK and $N_n = 7 \times 10^{13}$~cm$^{-3}$, the $^{209}$Pb($n,\gamma$) reaction rate $\lambda_{n}$ is equal to the $\beta$-decay rate $\lambda_{\beta}$. Above this $N_n$ value, $\lambda_{ n} > \lambda_{\rm \beta}$ and $^{210}$Pb can be significantly produced.  
Another way to break the cycle is to activate $^{210}$Bi($n,\gamma$), for which $\lambda_{ n} = \lambda_{\rm \beta}$ at $N_n = 2 \times 10^{12}$~cm$^{-3}$ (at $T=250$~MK) but then $^{211}$Bi experiences a quick $\alpha$-decay with $t_{1/2} = 128$~s. A neutron density of $N_n = 10^{15}$~cm$^{-3}$ is required to obtain $\lambda_{ n} = \lambda_{\rm \beta}$ at $^{211}$Bi.
To synthesize actinides, not only the s-process cycle needs to be broken but the nucleosynthesis path also needs to avoid $\beta$-decaying to $^{211-216}$Po isotopes that experience fast $\alpha$-decays ($t_{1/2}<1$~sec). This means that the neutron flux has to be large enough to produce $^{216-217}$Pb (or $^{217-218}$Bi) isotopes before they experience significant $\beta$-decays.
On the way to $^{216}$Pb, several Pb isotopes have half-lives requiring neutron densities above $10^{13} - 10^{15}$~cm$^{-3}$ to be bypassed (at $T=250$~MK). 
These simple considerations show that actinides may only produced if the neutron densities goes above typically $10^{14} - 10^{15}$~cm$^{-3}$, assuming the neutron capture rates can be accurately predicted by present reaction models, such as the TALYS code \citep{Koning23}.

\subsubsection{One zone calculations}

In a stellar model undergoing a PIE, convection mixes multiple zones with varying temperatures, densities, chemical compositions, and irradiation levels, resulting in a diluted production of elements. 
This effect cannot be captured by one-zone calculations, which may therefore yield different results compared to multizone stellar models (e.g. \cite{ingeberg25}). 
Although these simplified models must be interpreted with caution, they remain invaluable for carrying out rapid yet instructive nucleosynthesis calculations.
Here we use one-zone models to determine the neutron density threshold beyond which actinides begin to be produced.

We considered typical i-process conditions ($T=220$~MK and $\rho = 2500$~g~cm$^{-3}$). The initial composition was extracted from the TP of a full 1\Msun\ AGB model at [Fe/H]~$=-2.5$ just before the start of a PIE. Initial abundances of nuclei heavier than $^{56}$Fe were set to zero. We varied the initial proton mass fraction $X_{\rm H, ini}$ from $10^{-5}$ to $10^{-3}$ to mimic PIEs of different strengths (10 models in total). Within this range of proton mass fractions we obtained maximal neutron densities $ 7.9 \times 10^{13} < N_{\rm n, max} < 1.1 \times 10^{15}$~cm$^{-3}$ (Fig.~\ref{fig:onezone_nn}). Fig.~\ref{fig:onezone} shows the final mass fraction (after $6 \times 10^{5}$~s) of six representative elements. 
For $X_{\rm H, ini} < 3 \times 10^{-5}$, only the first peak s-process elements (e.g. Sr) are produced. At higher $X_{\rm H, ini}$, elements of the second (e.g. Ba) and eventually third peak (e.g. Pb) are synthesized. 
The production of Th and U (correlated with the production of Pb) starts to be non-negligible above $X_{\rm H, ini} \sim 2 \times 10^{-4}$ (corresponding to $N_{\rm n,max} \sim 5 \times 10^{14}$~cm$^{-3}$). 

\subsubsection{AGB models}\label{sec:agb}

In a full AGB model experiencing a PIE, the neutron density reaches $ 10^{15}$~cm$^{-3}$, which leads to the production of actinides \cite{choplin22b}. In our original work \cite{choplin22b}, three $(n,\gamma)$ reactions were missing in the nuclear network: $^{240}$Pa($n,\gamma$), $^{240}$U($n,\gamma$) and $^{240}$Np($n,\gamma$). These reactions are now included and the calculations performed again. We find that the resulting abundances were barely impacted. Fig.~\ref{fig:thu} shows the corrected version of the Fig.~1 published in \cite{choplin22b}. The only noticeable difference is the different black path around the U, Np, Pu region.

Figure~\ref{fig:actiyields1} and \ref{fig:actiyields2} show the yields in mass fraction\footnote{The yield $\mathcal{Y}_i$ of a nucleus $i$ is computed according to the relation $\mathcal{Y}_i =  \int_{0}^{\tau_{\rm star}} \dot{M}(t) \, X_{\rm i,s}(t) \, \text{d}t$ where $\tau_{\rm star}$ is total lifetime of the model star, and $X_{\rm i,s}(t)$ and $\dot{M}(t)$ are the surface mass fraction of nucleus, $i,$ and the mass-loss rate at time, $t$, respectively. The mass fraction yield is defined as $\mathcal{Y}_i / M_{\rm ej}$ with $M_{\rm ej} = \int_{0}^{\tau_{\rm star}} \dot{M}(t) \, \text{d}t$ the total mass ejected by the star.} of four SLR or long-lived isotopes: $^{232}$Th ($t_{1/2} = 14$~Gyr), 
$^{238}$U ($t_{1/2} = 4.46$~Gyr), $^{244}$Pu ($t_{1/2} = 81.3$~Myr) and $^{247}$Cm ($t_{1/2} = 15.6$~Myr).
In our models, the initial abundances of actinides are set to zero except for $^{232}$Th, $^{235}$U and $^{238}$U, where we take the present-day solar abundances \cite{asplund09, lodders03}, scaled to the model metallicity. 
The oblique dotted line in Fig.~\ref{fig:actiyields1} and \ref{fig:actiyields2} shows the initial actinides mass fraction in our models, which can be written as $X_{\rm ini}^{\rm ACT} = (X^{\rm 232Th}_{\odot} + X^{\rm 235U}_{\odot} + X^{\rm 238U}_{\odot}) \times 10^{\mathrm{[Fe/H]}}$. 
The models that produce actinides, i.e. with final actinide yields (in mass fraction) greater than $X_{\rm ini}^{\rm ACT}$ (shown by large symbols in Fig.~\ref{fig:actiyields1} and \ref{fig:actiyields2}) have [Fe/H]~$<-2.0$. Despite their low-metallicity, their $^{232}$Th and $^{238}$U yields can sometimes exceed the solar value (Fig.~\ref{fig:actiyields1}). At higher metallicities, the neutron-to-seed ratio is too low to synthesize actinides from Pb and Bi. In these AGB models however, neutron captures on existing actinides ($^{232}$Th, $^{235}$U and $^{238}$U) can change their distribution. 
Including overshooting in the calculations leads to shorter PIEs and smaller neutron exposures \cite{choplin24a}, which limits the production of actinides (Fig.~\ref{fig:actiyields1} and \ref{fig:actiyields2}). 

The production of actinides during PIEs in AGB stars is subject to large nuclear uncertainties  affecting the radiative neutron capture rates along the nuclear path, hence the Th and U abundances by up to $\sim 2-3$~dex \cite{martinet24}. 
We quantified the nuclear uncertainties affecting $^{232}$Th, $^{238}$U, $^{244}$Pu and $^{247}$Cm following the methodology of \citet{martinet24}, where we considered the Backward-Forward Monte Carlo (BFMC) approach, which provides a systematic framework to quantify and propagate parameter uncertainties in nuclear models by combining statistical sampling with experimental constraints.
In order to propagate nuclear uncertainties, 50 AGB models of 1\Msun\ AGB stars with [Fe/H]~$=-2.5$ were generated, applying random combinations of maximum and minimum neutron capture rates obtained with the BFMC method. 
To account for numerical uncertainties, only the models with final abundances (e.g. $^{232}$Th) within the 5$^{\rm th}$ to 95$^{\rm th}$ percentiles were retained (we refer to \cite{martinet24} for more details.)
The final uncertainties for these four isotopes, typically around 3 dex, are shown in Fig.~\ref{fig:actiyields1} and \ref{fig:actiyields2}.

\begin{table}[t]
\caption{Neutron density (in cm$^{-3}$) for which the $\beta$-decay and neutron capture rates of the (unstable) reference nucleus are equal ($\lambda_{\beta} = \lambda_{n}$) at 100~MK (typical radiative s-process conditions, third column) and 250~MK (typical i-process conditions, fourth column). 
Reference nuclei are isotopes where a critical branching is present on the nucleosynthesis path to the target nucleus. 
For a given reference nucleus, the table provides (at $T=100$~MK and 250~MK) the neutron density required to significantly produce the target.
The temperature-dependent $\beta$-decay rates are taken from \citet{takahashi87}. 
}\label{tab1} 
\begin{tabular}{@{}llll@{}}
\toprule
Reference & Target SLR& $T = 100$~MK  & $T = 250$~MK \\
\midrule
$^{59}$Fe    & $^{60}$Fe   & $3.7 \times 10^{10}$   & $4.8 \times 10^{10}$   \\
$^{125}$Sn    & $^{126}$Sn    & $8.4 \times 10^{11}$   & $6.5 \times 10^{12}$   \\
$^{128}$Sb    & $^{129}$I   & $4.7 \times 10^{11}$   & $1.5 \times 10^{12}$   \\
$^{127}$Te    & $^{129}$I   & $2.9 \times 10^{11}$   & $2.9 \times 10^{11}$   \\
$^{128}$I    & $^{129}$I   & $1.8 \times 10^{12}$   & $1.3 \times 10^{12}$   \\
$^{133}$Xe    & $^{135}$Cs   & $6.9 \times 10^{10}$   & $1.1 \times 10^{11}$   \\
$^{134}$Cs    & $^{135}$Cs   & $1.6 \times 10^{8}$   & $6.8 \times 10^{9}$   \\
$^{181}$Hf    & $^{182}$Hf   & $1.3 \times 10^{10}$   & $2.5 \times 10^{11}$   \\
\botrule
\end{tabular}
%\footnotetext{Source: This is an example of table footnote. This is an example of table footnote.}
%\footnotetext[1]{Example for a first table footnote. This is an example of table footnote.}
%\footnotetext[2]{Example for a second table footnote. This is an example of table footnote.}
\end{table}

\subsection{Production of other radionuclides}

We investigate here the synthesis of six SLR ($^{60}$Fe, $^{107}$Pd, $^{126}$Sn, $^{129}$I, $^{135}$Cs and $^{182}$Hf) in AGB models experiencing PIEs. 

\subsubsection{$^{60}\mathrm{Fe}$}\label{sec:fe60}

The $^{60}$Fe nuclide ($t_{1/2} = 2.62$~Myr in terrestrial conditions) can be synthesized from neutron capture on $^{59}$Fe which has a terrestrial half-life of 44.5~days. At $T = 100$~MK and 250~MK, $\lambda_n = \lambda_\beta$ is achieved for $N_n = 3.7 \times 10^{10}$ and $4.8 \times 10^{10}$~cm$^{-3}$, respectively (Table~\ref{tab1}). 
While the radiative s-process in AGB stars cannot reach such neutron densities, this is achievable via the reaction $^{22}$Ne($\alpha,n$) in the convective TP of intermediate mass AGB stars ($M_\ini \gtrsim 3-4$\Msun).
AGB stars models with $7 \lesssim M_\ini \lesssim 10$\Msun~have been shown to synthesize $^{60}$Fe \cite{trigo09,lugaro12a,doherty14, wasserburg17}.
Massive stars also produce $^{60}$Fe in their helium- and carbon-burning shells, both during hydrostatic and explosive stages \cite{timmes95, limongi06}. 
Other possible sites are electron-capture supernovae \cite{wanajo13} and type Ia supernovae \cite{woosley97}. 

Our AGB models experiencing PIEs reach maximal neutron densities of $6.8 \times 10^{13}< N_{\rm n, max} < 2.2 \times 10^{15}$~cm$^{-3}$, i.e. largely above the neutron density required to bypass $^{59}$Fe. 
We note that $^{60}$Fe can also be destroyed by neutron captures.
Overall, $^{60}$Fe is significantly produced (Fig.~\ref{fig:radio}, top left panel) and its yield scales with metallicity because of the increasing initial abundance of $^{56}$Fe.
%We quantified the nuclear uncertainties affecting $^{60}$Fe following the methodology of \citet{martinet24}, where we considered the Backward-Forward Monte Carlo (BFMC) approach, which provides a systematic framework to quantify and propagate parameter uncertainties in nuclear models by combining statistical sampling with experimental constraints.
%In order to propagate nuclear uncertainties, 50 AGB models of 1\Msun\ AGB stars with [Fe/H]~$=-2.5$ were generated, applying random combinations of maximum and minimum neutron capture rates obtained with the BFMC method. 
%To account for numerical uncertainties, only the models with final $^{60}$Fe abundances within the 5$^{\rm th}$ to 95$^{\rm th}$ percentiles were retained.
As for actinides (cf. Sect.~\ref{sec:agb}), we estimated the nuclear parameter uncertainties on the $^{60}$Fe yield based on \cite{martinet24}. We found a variation of 0.27~dex. 
%of the for $^{60}$Fe, the production of these SLRs are impacted by nuclear uncertainties. Based on \citet{martinet24}, we found variations of 0.41 and 0.81~dex for $^{126}$Sn and $^{129}$I, respectively.
%The final uncertainty on the $^{60}$Fe yield was found to be of 0.27~dex.  

We also indicated in the top left panel of Fig.~\ref{fig:radio} the range of $^{60}$Fe yields obtained in the $3 \leq M_\ini/\msun \leq 8$ AGB models computed in \citet{karakas10} for various metallicities. None of these models experience PIEs but $^{60}$Fe is synthesized in the TPs thanks to the convective s-process (see the discussion above). Their yields are higher than ours by $\sim 2$~dex at [Fe/H]~$=-2$ but comparable at solar metallicity. 
Whether AGB stars experiencing PIEs are important producers of $^{60}$Fe throughout cosmic history remains to be checked with galactic chemical evolution models.

\begin{figure}[t]
\centering
\includegraphics[width=0.49\textwidth]{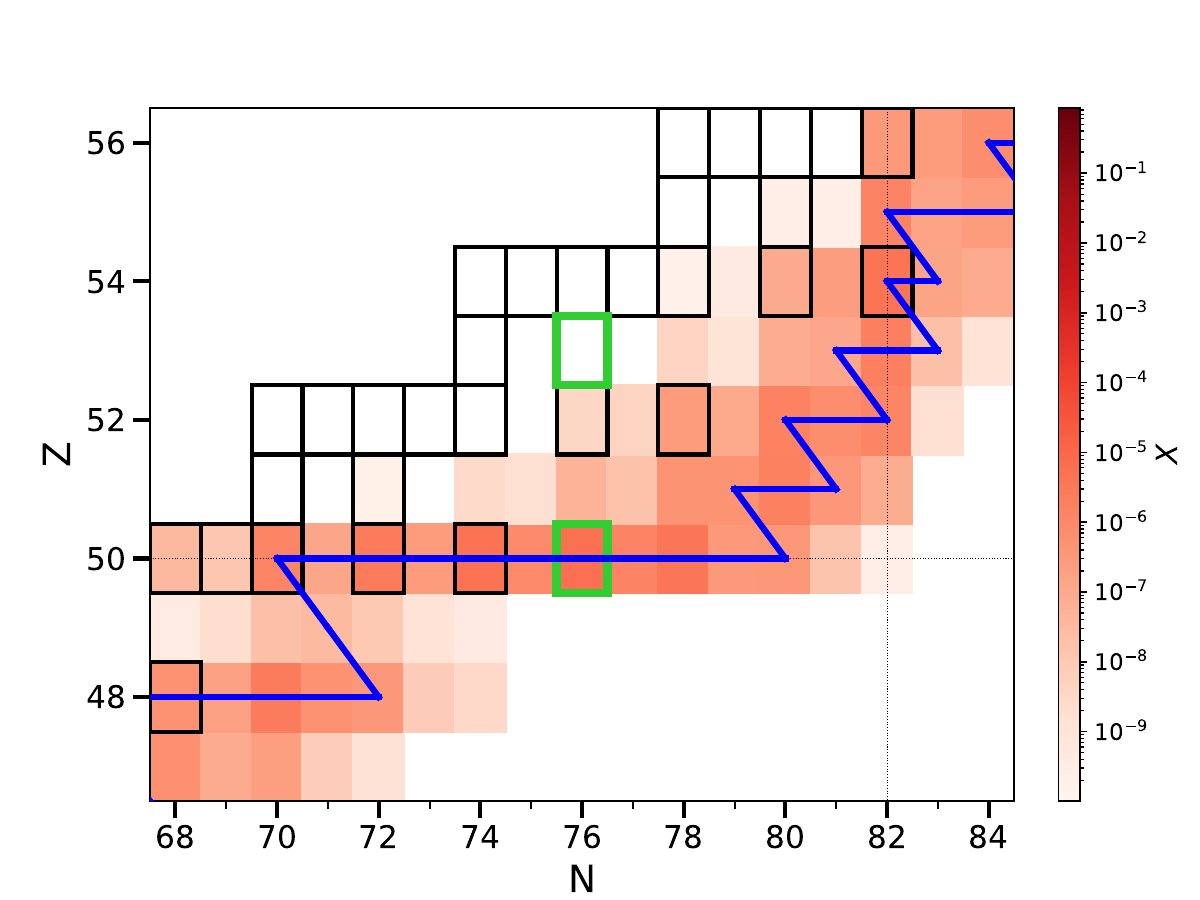}
\includegraphics[width=0.49\textwidth]{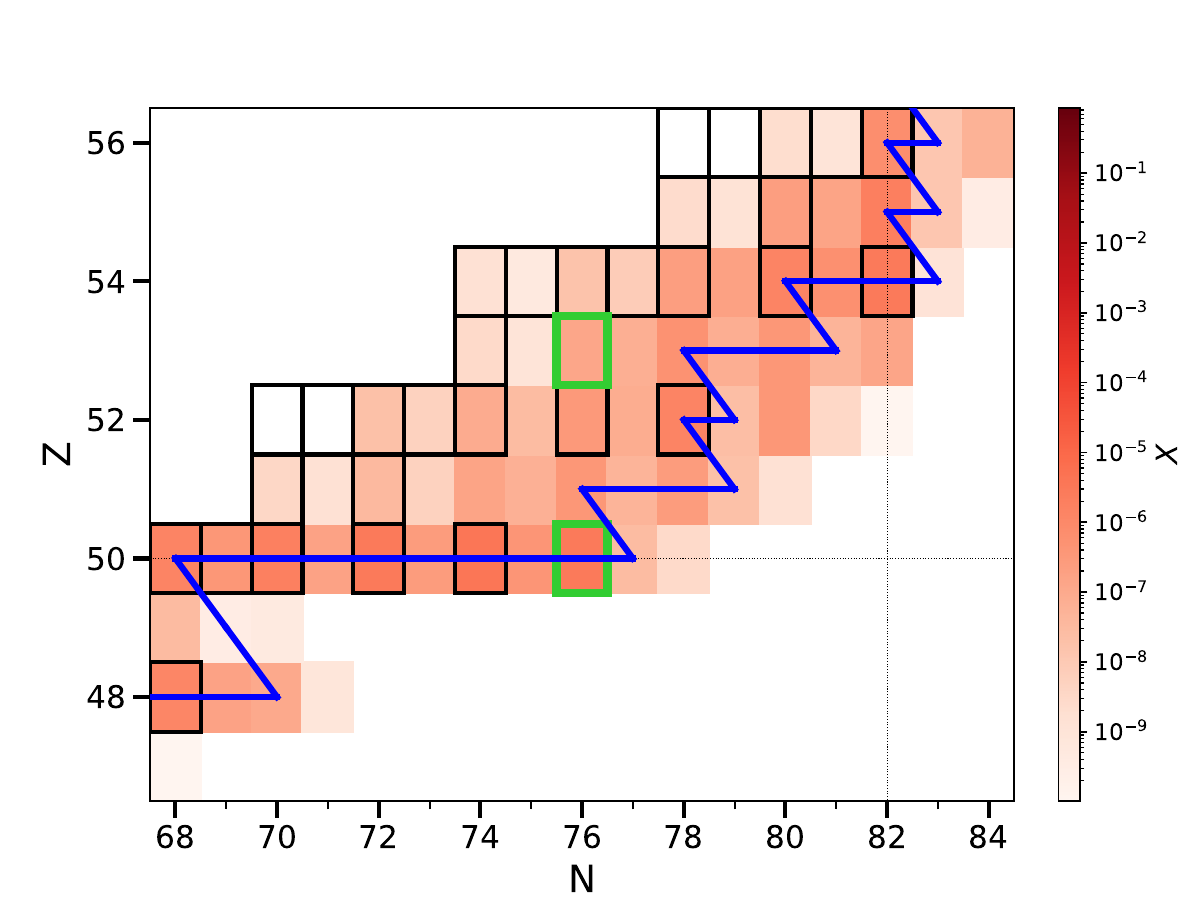}
\caption{Main i-process path (in blue) during the PIE of the 1\Msun\ AGB model at [Fe/H]~$-1.2$. 
Left panel: bottom of the TP, at maximal neutron density ($N_{\rm n} = 5.4 \times 10^{14}$~cm$^{-3}$). Right panel: at the same time but at a different location in the TP, where $N_{\rm n} = 10^{13}$~cm$^{-3}$. 
The black squares highlight the stable and long-lived nuclei. 
The two green squares shown the SLR $^{126}$Sn ($Z=50$) and $^{129}$I ($Z=53$).
The corresponding abundances of the models are shown by the red colour scale in mass fraction.
}
\label{fig:flux}
\end{figure}

\begin{figure}[h!]
\centering
\includegraphics[width=0.49\textwidth]{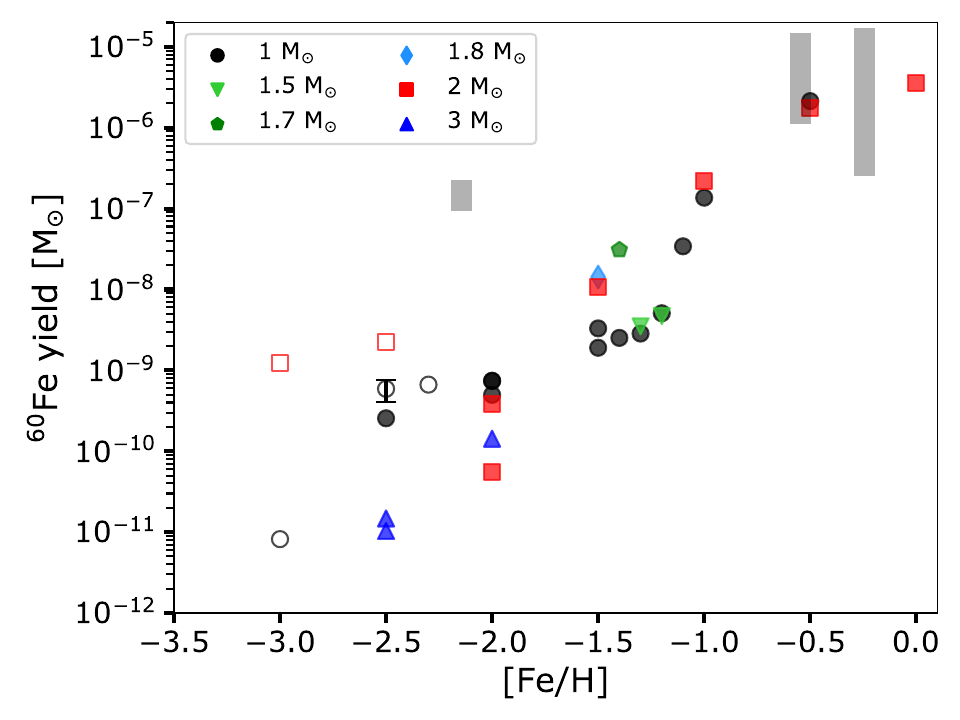}
\includegraphics[width=0.49\textwidth]{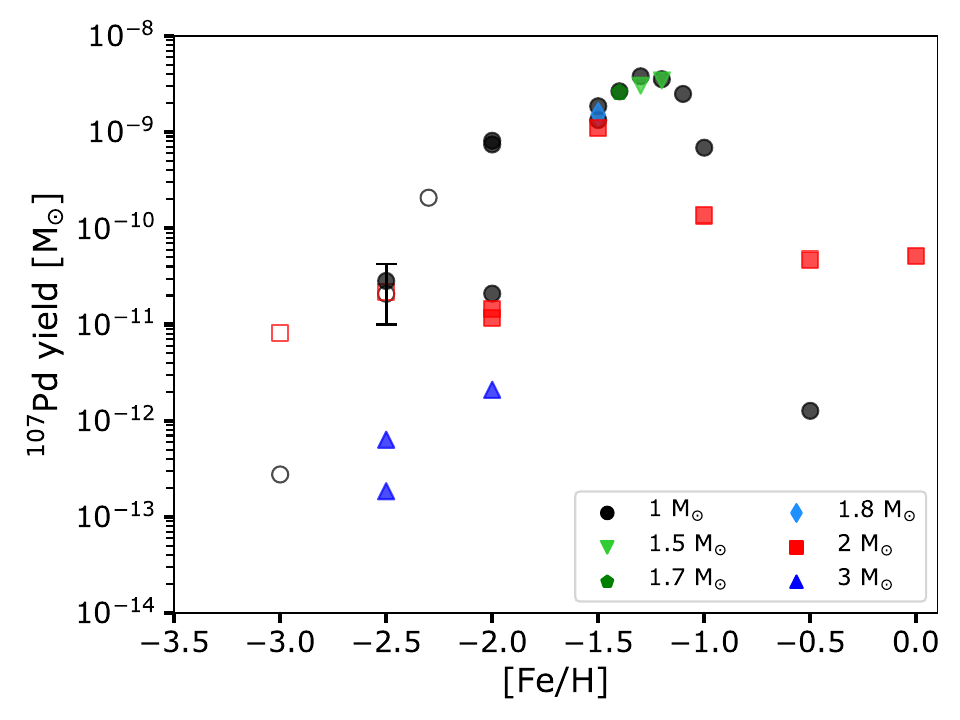}
\includegraphics[width=0.49\textwidth]{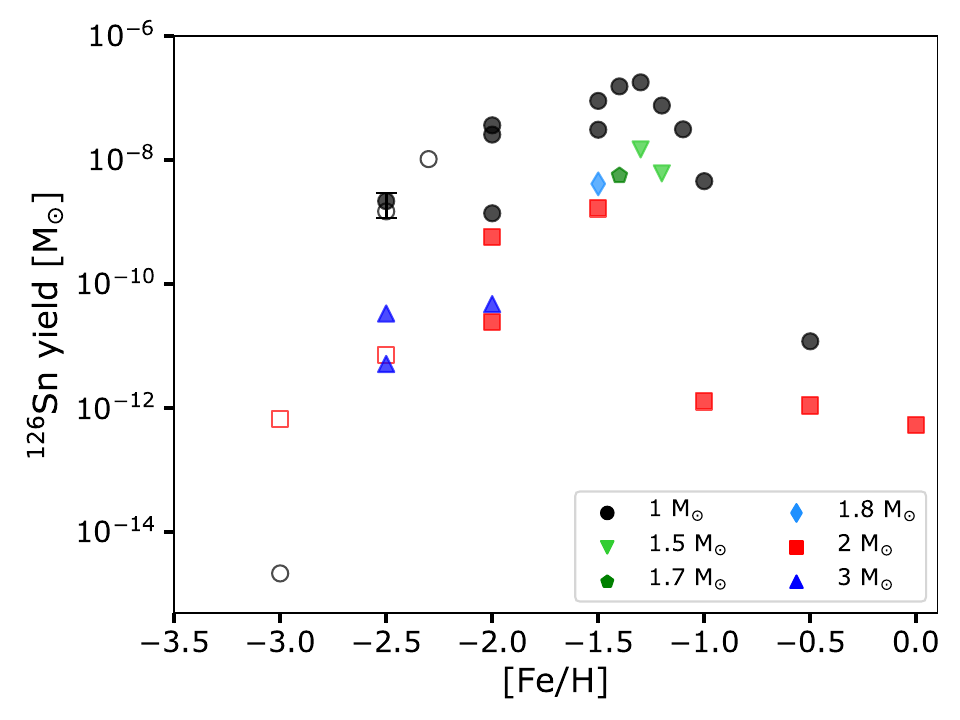}
\includegraphics[width=0.49\textwidth]{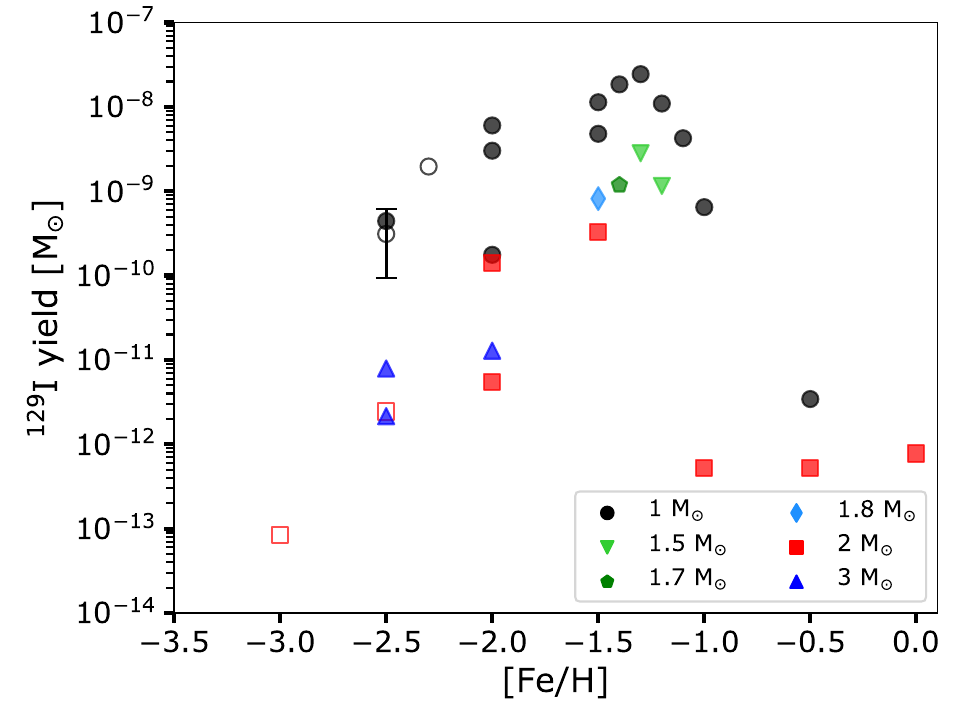}
\includegraphics[width=0.49\textwidth]{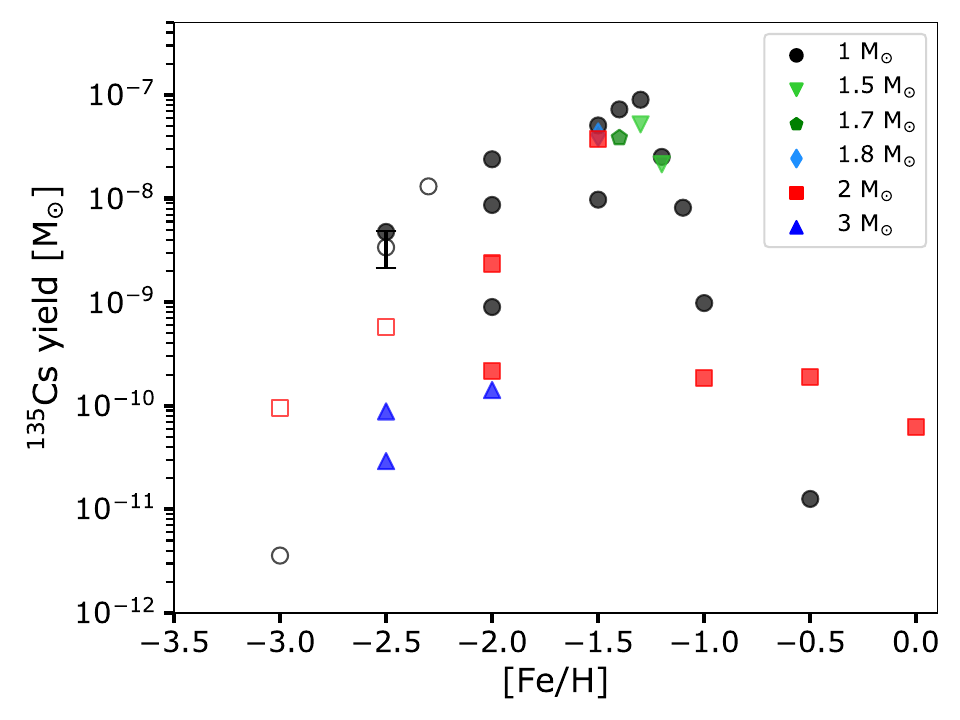}
\includegraphics[width=0.49\textwidth]{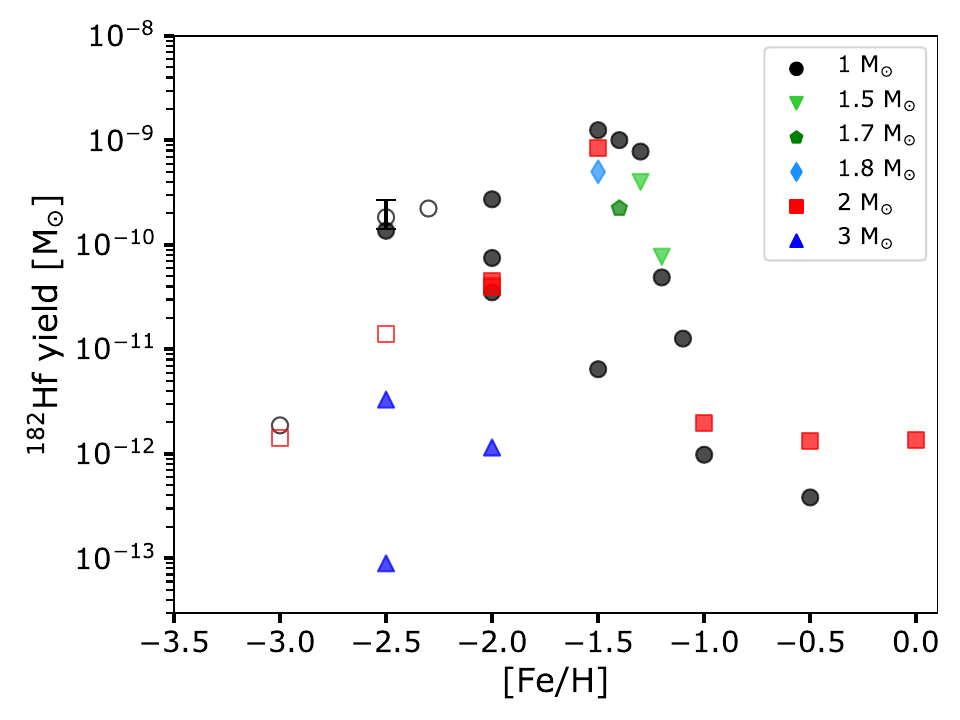}
\caption{Yields of six SLR from AGB models with different initial masses that experienced PIEs, as a function of metallicity. Open (filled) symbols represent models computed without (with) overshooting. Filled symbols with identical shape at a given metallicity correspond to models of same initial mass computed with different overshooting strengths $f_{\rm top}$ (cf. Sect.~\ref{sec2}). The error bars show nuclear parameter uncertainties associated with our [Fe/H]~$=-2.5$, 1~$M_{\odot}$ AGB model without overshoot, computed in \cite{martinet24} (see Sect.~\ref{sec:agb} for details). In the top left panel, the three shaded area represent the range of $^{60}$Fe yield for the $3 \leq M_\ini/\msun \leq 8$ AGB models (not experiencing PIEs) at metallicities $Z=0.0001$ ([Fe/H]~$=-2.1$), $Z=0.004$ ([Fe/H]~$=-0.5$) and $Z=0.008$ ([Fe/H]~$=-0.2$) published in \citet{karakas10}.}
\label{fig:radio}
\end{figure}

\begin{figure}[t]
\centering
\includegraphics[width=0.49\textwidth]{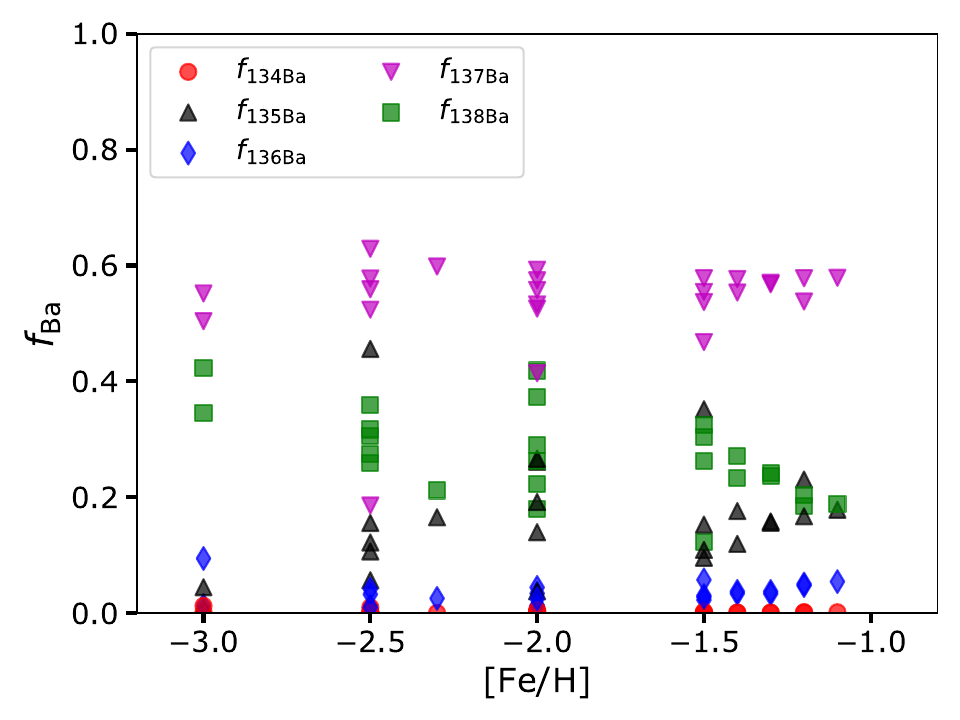}
\includegraphics[width=0.49\textwidth]{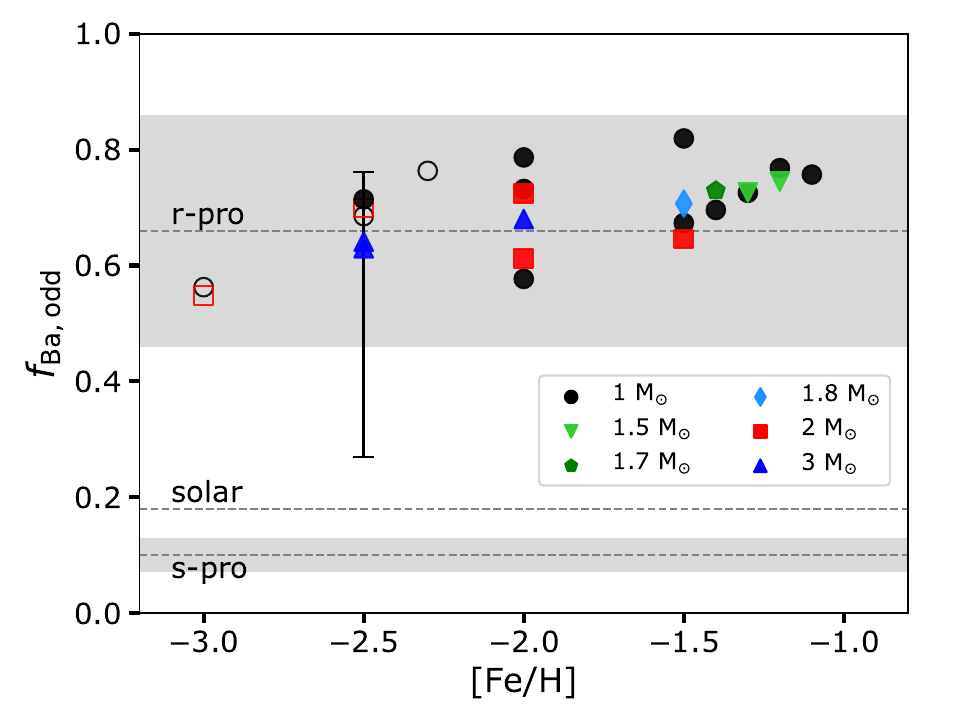}
\caption{ Left panel: fraction of the five stable barium isotopes $f_{\rm Ba}$ in the yields of our AGB models that experienced PIEs, after the decay of the unstable isotopes (including the SLR $^{135}$Cs). Right panel: contribution from odd Ba isotopes $f_{\rm Ba, odd} = f_{\rm 135Ba} + f_{\rm 137Ba}$. 
Models with [Fe/H]~$\geq -1$ are not shown because they do not produce barium significantly (see text for details).
Dashed lines indicate $f_{\rm Ba, odd}$ in the Sun, in the solar r-process \citep{Beer97,goriely99a, arlandini99, Prantzos20}, and resulting from s-process nucleosynthesis \citep{goriely18c}. Shaded rectangles represent the uncertainties associated to $f_{\rm Ba, odd}$ for the r-process and s-process. 
Nuclear uncertainties associated to the 1\Msun\ model at [Fe/H]~$=-2.5$ are represented by the error bar (see text for details). Open (filled) symbols represent models computed without (with) overshooting. Filled symbols with identical shape at a given metallicity correspond to models of same initial mass computed with different overshooting strengths $f_{\rm top}$ (cf. Sect.~\ref{sec2}).
}
\label{fig:baf}
\end{figure}

\subsubsection{$^{126}\mathrm{Sn}$ and $^{129}\mathrm{I}$}

The $^{126}$Sn SLR ($t_{1/2} = 0.2$~Myr) can be synthesized if the unstable $^{121}$Sn, $^{123}$Sn and $^{125}$Sn nuclei are bypassed ($^{122}$Sn and $^{124}$Sn are stable). Neutron densities of $N_n \simeq 10^{10} - 10^{11}$~cm$^{-3}$ are required to obtain $\lambda_\beta = \lambda_n$ at $^{121}$Sn and $^{123}$Sn. The $^{125}$Sn isotope requires the highest neutron density with $N_n \simeq 10^{12} - 10^{13}$~cm$^{-3}$ to reach $\lambda_\beta = \lambda_n$ (Table~\ref{tab1}). 
Because of the high neutron density needed to synthesize $^{126}$Sn, the production of this SLR has usually been associated with the r-process nucleosynthesis \cite{lugaro18}.

During a PIE, $N_n$ can exceed $10^{12}$~cm$^{-3}$ by several orders of magnitude, allowing the production of $^{126}$Sn (Fig.~\ref{fig:flux} and \ref{fig:radio}). 
The production of $^{126}$Sn peaks around [Fe/H]~$=-1.3$ (Fig.~\ref{fig:radio}). At lower metallicities, the higher neutron-to-seed ratio (with the seeds mostly under the form of $^{56}$Fe) favors the synthesis of heavier elements to the detriment of $^{126}$Sn. At higher metallicities, because of the larger initial abundance of $^{56}$Fe (and therefore the lower neutron-to-seed ratio), the synthesis of heavier elements is less important.

The $^{129}$I isotope ($t_{1/2} = 16.14$~Myr in terrestrial conditions) can be synthesized from $^{127}$I($n,\gamma$)$^{128}$I($n,\gamma$)$^{129}$I, from the chain $^{126}$Te($n,\gamma$)$^{127}$Te($n,\gamma$)$^{128}$Te($n,\gamma$)$^{129}$Te($\beta^-$)$^{129}$I or from neutron captures within the Sb isotopic chain up to $^{129}$Sb followed by two $\beta^-$-decays. During these sequences, the $^{128}$Sb, $^{127}$Te and $^{128}$I nuclei can be bypassed ($\lambda_n = \lambda_\beta$) for neutron densities of $10^{11} - 10^{12}$~cm$^{-3}$ (Table~\ref{tab1}). At higher neutron densities, $^{129}$I can be synthesized by other channels (e.g. via the decay of $^{129}$Sn).
As for $^{126}$Sn, the production of the SLR $^{129}$I has usually been associated with the r-process nucleosynthesis \cite{meyer00b, lugaro18}. 

In our i-process AGB models experiencing PIEs, $^{129}$I is synthesized by different channels inside the TP. At the bottom, where the neutron density is the highest, $^{129}$I mainly comes from the decay of $^{129}$Sn and $^{129}$Sb (Fig.~\ref{fig:flux}, left panel). Higher up in the pulse, where the neutron density is lower, $^{129}$Sn is not synthesized and $^{129}$I mostly originates from the decay of $^{129}$Sb and $^{129}$Te (Fig.~\ref{fig:flux}, right panel).
In the end, $^{129}$I is substantially synthesized in AGB stars experiencing PIEs  (Fig.~\ref{fig:radio}). The $^{129}$I yields follows a similar trend with metallicity as $^{126}$Sn, with a production peak at [Fe/H]~$=-1.3$. 
As for $^{60}$Fe, the production of these SLRs are impacted by nuclear uncertainties. Based on \citet{martinet24}, we found variations of 0.41 and 0.81~dex for $^{126}$Sn and $^{129}$I, respectively.

\subsubsection{$^{107}\mathrm{Pd}$, $^{135}\mathrm{Cs}$ and $^{182}\mathrm{Hf}$}

$^{107}$Pd can be synthesized through neutron captures on the stable $^{106}$Pd nuclide by the s-process. Similarly,  
$^{135}$Cs comes from $^{133}$Cs($n,\gamma$)$^{134}$Cs($n,\gamma$). The $^{134}$Cs $\beta^-$ half-life varies by more than 3 orders of magnitude from 100~MK to 250~MK, \cite{takahashi87, li21} and requires $N_n \simeq 10^8 - 10^{10}$~cm$^{-3}$ to be bypassed (Table~\ref{tab1}). 
The situation is similar for $^{182}$Hf, which requires $N_n \simeq 10^{10} - 10^{11}$~cm$^{-3}$ to have $\lambda_n = \lambda_\beta$ at $^{181}$Hf.

Contrary to $^{129}$I and $^{126}$Sn, these three isotopes can be synthesized in AGB stars during radiative and/or convective s-process nucleosynthesis \cite{trigo09, lugaro14, bisterzo15, wasserburg17, li21}. 
They are also produced in AGB models experiencing i-process during PIEs (Fig.~\ref{fig:radio}). The yield dependence with metallicity is similar to $^{129}$I and $^{126}$Sn, with a production peak at [Fe/H]~$= -1.3$ ($-1.5$ for $^{182}$Hf). 
The impact of nuclear uncertainties based on \citet{martinet24} are 0.63, 0.35 and 0.28~dex for $^{107}$Pd, $^{135}$Cs and $^{182}$Hf, respectively.

\subsection{Barium isotopic ratios}

The determination of specific isotopic abundance ratios can set constraints on the nucleosynthesis process at their origin. For this purpose, Ba isotopes are of particular interest since the spectral lines of odd-$N$ isotopes are affected by hyperfine splitting \cite{Vaneck24}.

The left panel of Fig.~\ref{fig:baf} shows the fraction of stable barium isotopes in our AGB models, expressed as $Y_{\mathrm{Ba},i} / \sum_{i} Y_{\mathrm{Ba},i}$ where $Y_{\mathrm{Ba},i} = X_{\mathrm{Ba},i} / A_i$ is the number abundance of the considered barium isotope with $X_{\mathrm{Ba},i}$ its mass fraction and $A_i$ its atomic mass. 
Our AGB models experiencing PIEs with [Fe/H]~$\geq -1$ do not produce Ba (and heavier elements) significantly  (i.e. [Ba/Fe]~$\simeq 0$) because of the too low neutron-to-seed ratio (see Fig.~4 in \citet{choplin24b}).
At [Fe/H]~$<-1$, $^{137}$Ba is the dominant isotope ($\sim 60$~\%) followed by $^{138}$Ba ($\simeq 20 - 40$~\%). $^{137}$Ba is mostly synthesized out of the decay of $^{137}$Cs and $^{137}$Xe (Fig.~\ref{fig:flux}).

The resulting $f_{\rm Ba, odd} = f_{\rm 135Ba} + f_{\rm 137Ba}$, of particular interest in spectroscopic observations, is shown in the right panel of Fig.~\ref{fig:baf}. At [Fe/H]~$<-1$, our $1 \leq M_\ini/\msun \leq 3$ PIE models computed with our recommended set of reaction rates lead to $f_{\rm Ba, odd} \simeq 0.6 - 0.8$ (i.e. similar to the r-process). 
As before, to quantify the nuclear uncertainties affecting $f_{\rm Ba,odd}$, we followed the methodology from \citet{martinet24}. 
We retained only the AGB models with final surface abundances within the 5$^{\rm th}$ to 95$^{\rm th}$ percentiles for each of the five stable Ba isotopes.
The 29 remaining models predict $0.27<f_{\rm Ba, odd}<0.76$. 
In the end, our reference 1\Msun\ AGB model with [Fe/H]~$=-2.5$ has $f_{\rm Ba, odd} = 0.67^{+0.09}_{-0.40}$ (Fig.~\ref{fig:baf}, right panel) which is clearly above the s-process and solar values. 

An increasing number of stars with chemical abundance patterns intermediate between s- and r- processes are being observed (the r/s-stars, e.g. \citep{mishenina15, roederer16, karinkuzhi21,karinkuzhi23, mashonkina23, hansen23}). 
One possible scenario assumes that these stars were born in binary systems with a slightly more massive companion that evolved to the AGB phase, produced and eventually ejected i-process material through stellar winds.
Some of the wind was accreted by the lower-mass companion that became a r/s-star. During the accretion process, the AGB matter is diluted in the companion envelope, which can modify the value of $f_{rm Ba, odd}$. Nevertheless, we have verified that, in general, the dilution is not strong enough to significantly modify the value of $f_{\rm Ba, odd}$.

\section{Summary and conclusion}\label{concl}

In this paper, after a quick review of the three different modes of production of heavy elements in AGB stars, we investigated the production of actinides and of six SLR ($^{60}$Fe, $^{107}$Pd, $^{126}$Sn, $^{129}$I, $^{135}$Cs and $^{182}$Hf) during i-process nucleosynthesis in AGB stars models computed with the stellar evolution code \textsc{STAREVOL}, with initial masses $1 \leq M_\ini/\msun \leq 3$ and metallicities $-3 \leq $~[Fe/H]~$ \leq 0$. 

Actinides (particularly $^{232}$Th, $^{238}$U, $^{244}$Pu and $^{247}$Cm) are synthesized in AGB models with [Fe/H]~$<-2$ and their production is higher in models without overshooting mixing. 
The $^{60}$Fe isotope is synthesized at all metallicities and all initial masses. Its abundance in the yield scales with metallicity. 
The $^{126}$Sn and $^{129}$I SLR, which are generally attributed to r-process nucleosynthesis exclusively, are produced at all masses and metallicities, with a production peak at [Fe/H]~$=-1.3$. 
The synthesis of $^{107}$Pd, $^{135}$Cs and $^{182}$Hf follows the same  metallicity dependance as $^{126}$Sn and $^{129}$I.

While these radionuclides are generally attributed to the s- and/or r-process nucleosynthesis only, the present work shows that actinides and several SLR are synthesized during i-process nucleosynthesis in AGB stars, particularly at low-metallicity (except for $^{60}$Fe). 
Galactic chemical evolution models following SLR (e.g. \cite{cote19,wehmeyer24}) are required to assess the possible contribution of AGB stars having experienced i-process nucleosynthesis. 

Finally, we find a fraction of odd Ba isotopes, $f_{\rm Ba, odd} \simeq 0.6 - 0.8$ in our i-process AGB models with $1 \leq M_\ini/\msun \leq 3$ and [Fe/H]~$<-1$. At [Fe/H]~$=-2.5$, the nuclear uncertainties predict $0.27<f_{\rm Ba, odd}<0.76$, i.e. clearly above the s-process value of $0.10\pm 0.03$ for AGB stars.

\bmhead{Acknowledgements}

A.C. is a F.R.S-FNRS fellow. L.S. and S.G. are senior F.R.S-FNRS research associates. The authors are members of BLU-ULB, the interfaculty research group focusing on space research at ULB - Université libre de Bruxelles.

\bibliography{astro.bib}% common bib file

\end{document}